\documentclass{article}
\usepackage{spconf,amsmath,graphicx}

\usepackage{cite}
\usepackage{amssymb,amsfonts}
\usepackage{textcomp}
\usepackage{xcolor}
\usepackage{epsfig,bm,subfigure}
\usepackage{multirow}

\usepackage{algpseudocode}
\usepackage{tabu}
\usepackage{makecell}

\usepackage{threeparttable}
\usepackage{url}
\usepackage{booktabs,bigstrut}
\usepackage{setspace}

\begin{document}
\title{Fast QTMT Partition for VVC Intra Coding Using U-Net Framework}

\name{Zhao Zan$^{\star}$ \qquad Leilei Huang$^{\ddag}$ \qquad ShuShi Chen$^{\star}$ \qquad Xiantao Zhang$^{\ast}$}
\name{Zhenghui Zhao$^{\ast}$ \qquad Haibing Yin$^{\dag}$  \qquad Yibo Fan$^{\star}$ 
\thanks{This work was supported in part by the National Natural Science Foundation of China under Grant 62031009, in part by Alibaba Innovative Research (AIR) Program, in part by the Fudan University-CIOMP Joint Fund(FC2019-001), in part by the Fudan-ZTE Joint Lab, in part by Pioneering Project of Academy for Engineering and Technology Fudan University(gyy2021-001), in part by CCF-Alibaba Innovative Research Fund For Young Scholars. (Corresponding author:
Leilei Huang, llhuang@cee.ecnu.edu.cn)
}
}

\address{$^{\star}$Fudan University, Shanghai, China \qquad $^{\ddag}$East China Normal University, Shanghai, China \\
 $^{\ast}$Alibaba Group, Hangzhou, China \qquad $^{\dag}$Hangzhou Dianzi University, Hangzhou, China \qquad  }

%
\maketitle
\begin{abstract}
Versatile Video Coding (VVC) has significantly increased encoding efficiency at the expense of numerous complex coding tools, particularly the flexible Quad-Tree plus Multi-type Tree (QTMT) block partition. This paper proposes a deep learning-based algorithm applied in fast QTMT partition for VVC intra coding. Our solution greatly reduces encoding time by early termination of less-likely intra prediction and partitions with negligible BD-BR increase. Firstly, a redesigned U-Net is recommended as the network's fundamental framework. Next, we design a Quality Parameter (QP) fusion network to regulate the effect of QPs on the partition results. Finally, we adopt a refined post-processing strategy to better balance encoding performance and complexity. Experimental results demonstrate that our solution outperforms the state-of-the-art works with a complexity reduction of 44.74\% to 68.76\% and a BD-BR increase of 0.60\% to 2.33\%.
\end{abstract}

\section{Introduction}
\label{sec:intro}

\begin{figure}
  \centering
  \includegraphics[width=7.2cm,height=4.3cm]{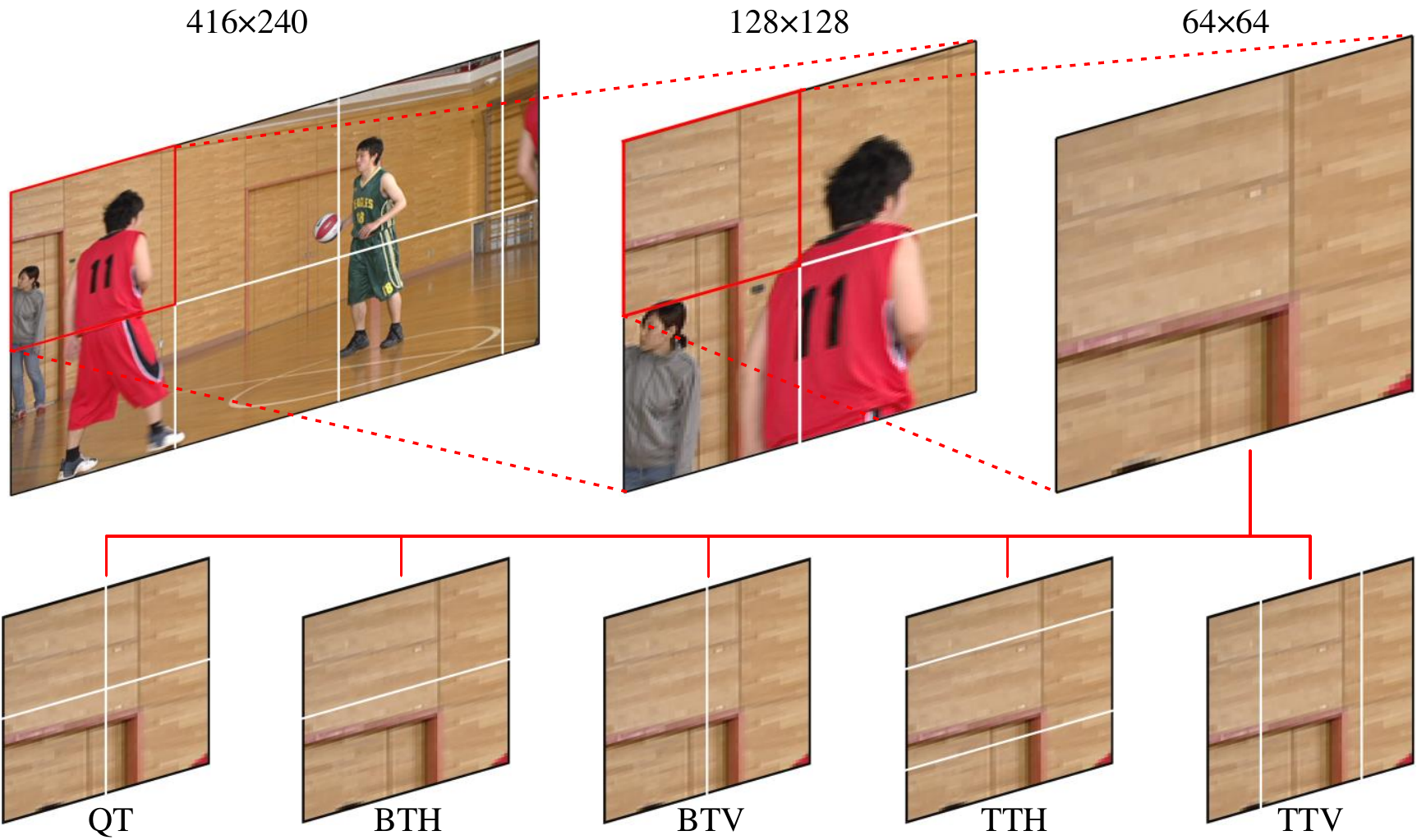}
  \caption{A particular example of QTMT partition in VTM.} 
  \label{QTMT}
\end{figure}

The expansion of applications and the explosive growth of data has brought about new challenges for video encoding. Against this background, VVC\cite{2021BrossVVC} was released. VVC applies QTMT technology\cite{huang2019QTMT}, which adds Binary Tree Horizontal/Vertical (BTH/BTV) and Ternary Tree Horizontal/Vertical (TTH/TTV) partitions in addition to the Quad-Tree (QT) partition that is already supported in High-Efficiency Video Coding (HEVC)\cite{2012SullivanHEVC}. Coding tree units with the size of 128×128 must first be executed with a QT partition for intra coding in VVC Test Model (VTM). Then the time-consuming Rate-Distortion Optimization (RDO) procedure determines the optimal partition structure, with a minimum sub-block size of 4. The complexity comes from two processes: the performing of intra prediction and the conduction of various partitions. Fig. \ref{QTMT} shows an example of QTMT partition.

Numerous studies \cite{Shen2014Effective, Min2015Fast, chen2020learned} have shed light on the fast partition for HEVC, but they cannot be directly applied to VVC. Fast partition algorithms for VVC are of increasing interest, and the literature may be roughly separated into two categories: one based on statistical analysis \cite{Fan2020Fast, yang2019low, dong2021fast, 2021TCSVTcfg, 2021ISCASSVM} and the other employing Convolutional Neural Networks (CNN) \cite{2020ICIPCNN, zhang2021fast, 2021TIPDeepQTMT, 2022TCSVTHGFCN}. 

\begin{figure*}[ht]
  \centering
  \includegraphics[width=16cm,height=8cm]{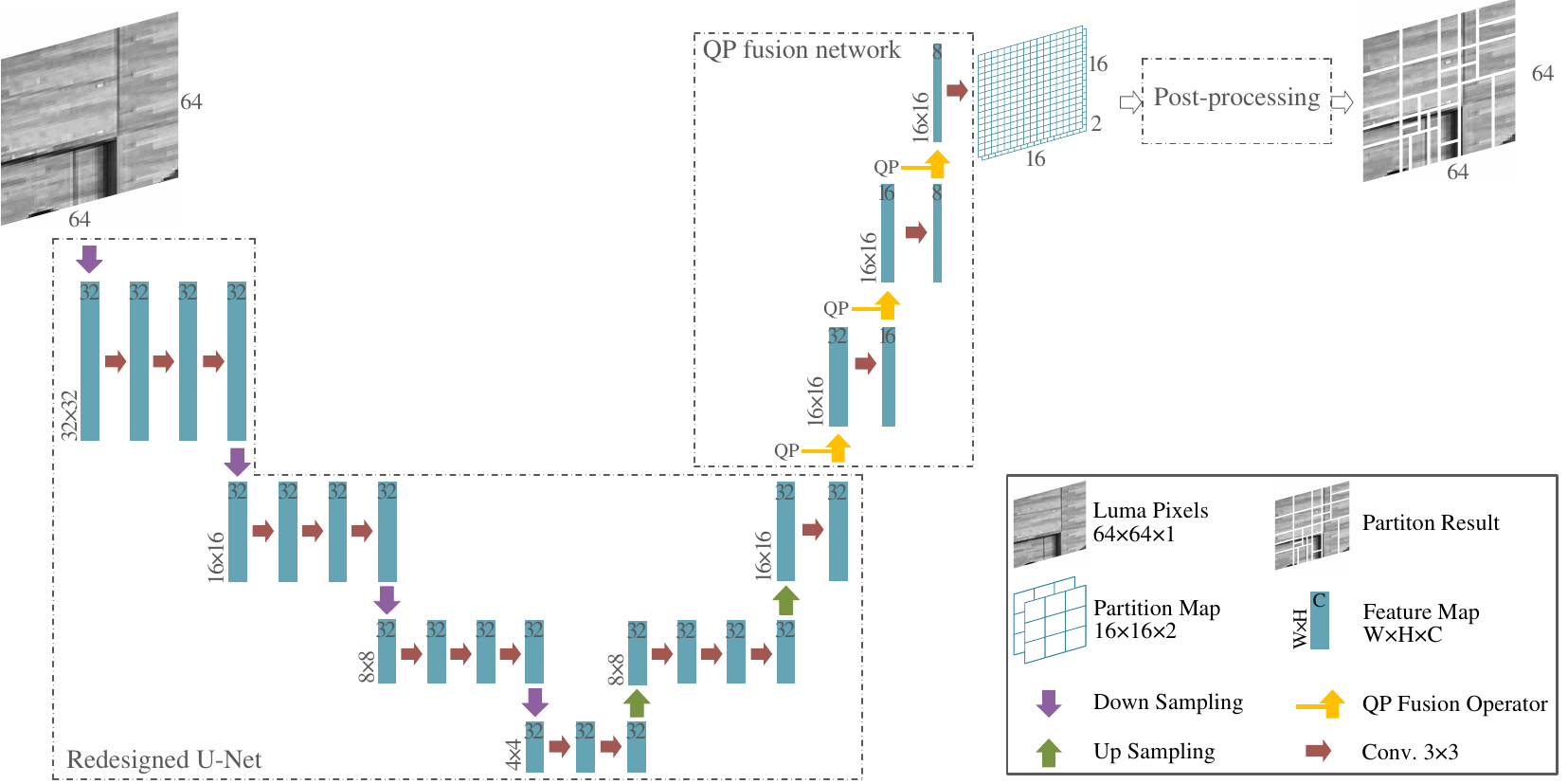}
  \caption{The overall framework. Dashed boxes indicate key components, including a redesigned U-Net, a QP fusion network, and a post-processing procedure.} \label{framework}
\end{figure*}

In the methods based on statistical analysis, useful features closely related to the partition results are calculated and fed into classifiers. The classifiers proposed in these methods are different. Fan \textit{et al.} \cite{Fan2020Fast} proposed a decision approach by simply comparing the feature values with given thresholds. Machine learning was also involved in the design of classifiers like decision tree \cite{yang2019low, dong2021fast}, LGBM \cite{2021TCSVTcfg}, and SVM \cite{2021ISCASSVM}. These classifiers are typically called recursively from the top down, making it impossible to process them in parallel. Moreover, if a specific feature is only available in VVC, such as the QTMT depth level \cite{2021TCSVTcfg}, this type of method cannot be directly migrated to other video coding standards.

CNN-based methods have also received much attention. An essentially similar approach was utilized to obtain partition probabilities of all 4×4 sub-blocks directly \cite{2020ICIPCNN, zhang2021fast}. This approach has no sequential data dependency, but its performance is mediocre due to the simple network design and the relatively rough post-processing strategy that does not consider whether to perform intra prediction. A multi-stage exit CNN model was proposed to decide whether to split the Coding Unit (CU) at each stage to reduce much encoding time \cite{2021TIPDeepQTMT}. This method is also subject to its high network complexity and lack of parallelism. Wu \textit{et al.} \cite{2022TCSVTHGFCN} proposed a fully convolutional network, striking a decent compromise between encoding time and quality using a hierarchical grid map. However, this work utilizes two distinct networks for CUs of size 64×64 and CUs of other sizes, which makes network training and deployment more challenging. Additionally, there is room for enhancement because the impact of QPs on the partition is not considered in the network.

This paper proposes a CNN-based fast partition algorithm for VVC intra coding. Our solution can have massive parallelism and hardware implementation because it merely takes the input of original pixels and QPs. Another attractive feature is the feasibility of being migrated to other video coding standards, as it does not require any intermediate coding information closely tied to VVC. In brief, Our specific contributions are summarized as follows: (1) a redesigned U-Net \cite{2015RonnebergerUNet} is employed as the network's fundamental framework, which extracts texture information effectively; (2) we propose a QP fusion network to regulate the effect of QPs on the partition results; and (3) a refined post-processing strategy is adopted to optimize the algorithm further and ultimately obtain higher performance beyond the state-of-the-art works.

\section{Proposed Method}
\label{sec:method}

\subsection{Overview}
Fig. \ref{framework} illustrates the proposed fast partition algorithm procedure. The framework comprises three key components, a redesigned U-Net, a QP fusion network, and a post-processing procedure. The former two take the original luma pixels and QP and output the partition probabilities of all 4×4 sub-blocks, called ``partition map". The latter takes this map and outputs the final QTMT partition.

\subsection{Redesigned U-Net Structure}
The block partition task is comparable to the traditional semantic segmentation task in image processing. Based on the information like content texture, the current image is split into non-overlapping sub-blocks in both tasks. However, the splitting edge of the block partition task is regular, while that of the semantic segmentation task can be irregular. Therefore, the U-Net, a classic network widely used for semantic segmentation, has been drastically redesigned to make it better suited for the block partition task. The redesigned U-Net can more fully extract and merge the features of different sub-blocks in 64×64 input blocks. 

The output is modified to be the partition map with a shape of 16×16×2, which contains the partition probability of 4×4 sub-blocks. The vertical and horizontal partition probabilities are recorded in two channels, respectively. Although the bottom row and right column of the partition map are redundant, the shape of 16×16×2 is selected to maintain the regularity of the feature maps, rather than two different shapes of 16×15 and 15×16.

As illustrated in Fig.\ref{framework}, the network structure also contains the following changes over the original U-Net: (1) decrease the number of convolutional kernels to lower the complexity of the network; (2) remove the bridge structure because it increases the complexity and has no discernible effect on final performance for the specific task; (3) use the CNN with a stride value of 2 for downsampling, and the sub-pixel CNN \cite{shi2016real} for upsampling; and (4) ensure the size of all feature maps is an integer multiple of 4 by using the ``same" padding.

\subsection{QP Fusion Network}
QPs have a significant impact on the partition. The role of QPs cannot be adequately explored if the regulation of QPs is just intermingled in the post-processing procedure. Consequently, the QP fusion network is added before the output of the redesigned U-Net. As shown in Fig.\ref{framework}, the QP fusion network consists of a cascade of three components, each containing a QP fusion operator and a convolutional layer. Half of the input feature map is first divided by the normalized QPs, and convolutions perform the feature fusion with a kernel size of 3×3. It has been demonstrated through ablation experiments that the three-layer cascade structure serves better than other QP fusion structures.

\begin{figure}[t]
  \centering
  \includegraphics[width=8.5cm,height=6.5cm]{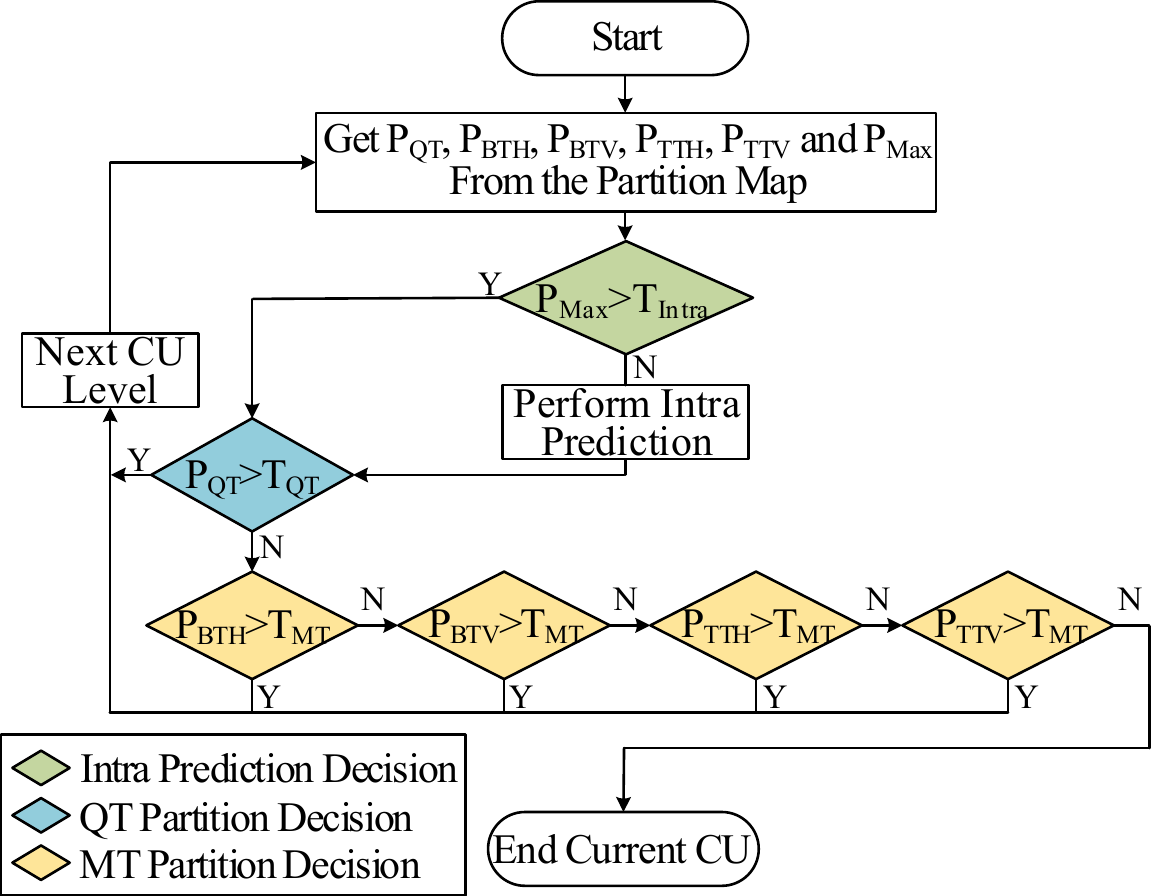}
  \caption{ The flowchart of the post-processing procedure.} \label{flowchart}
\end{figure}

\subsection{Refined Post-processing Strategy}
After the redesigned U-Net and the QP fusion network, a refined post-processing strategy is recommended. The post-processing procedure is shown in Fig.\ref{flowchart}. First, calculate the probability of each of the five partitions and the maximum value according to the partition map. Three different thresholds are proposed in total: the threshold $\rm T_{Intra}$ to decide whether to perform intra prediction and the thresholds $\rm T_{QT}$ and $\rm T_{MT}$ to determine whether to skip the QT partition and the MT partition, respectively. If $\rm P_{Max}$ is greater than $\rm T_{Intra}$, there is a high probability that some partition will be performed at this time, so the intra prediction is unnecessary. After that, if the likelihood of a specific partition is greater than the corresponding threshold, the partition is performed; otherwise, the RDO procedure is early terminated.

As the partition depth increases, the size of the CUs decreases, and fewer values in the partition map are used to calculate the partition probabilities, reducing the results' robustness. The threshold $\rm T_{QT}$ and $\rm T_{MT}$ should gradually decrease as the partition depth rises so that more potential partitions of smaller blocks can enter the regular RDO procedure, preventing a significant loss in encoding quality.

\section{Experiment}
\label{sec:Experiment}

\subsection{Experimental Setting} 
DIV2K \cite{2017AgustssonDIV2K} and RAISE \cite{2015DangRAISE} are utilized to create the training set. The training framework is TensorFlow 15.0 \cite{tensorflow}. Binary\_cross\_entropy is used as the loss function, and Adam \cite{kingma2015adam} is employed as the optimizer. The models are trained for 120 epochs, with an initial learning rate of 1e-3 which decreases by 20\% for every 10 epochs.

The proposed approach is implemented in the reference software VTM 7.0. The threshold settings in the post-processing are shown in Table \ref{Thres}. The test sequences include all 22 video sequences in Class A-E. The encoder\_intra\_vtm.cfg in VTM is used as the encoder configuration file with QPs of 22, 27, 32, and 37. BD-BR \cite{bjontegaard2001calculation} and complexity reduction $\rm \Delta$T are measured separately to evaluate the performance. The calculation of $\rm \Delta$T is shown below:
\begin{equation}\label{eq1}
  {\rm \Delta{T}}= \frac{1}{4}\sum_{i=1}^{4}\frac{T_{anchor}(i)-T_{test}(i)}{T_{anchor}(i)}
\end{equation}
where $T_{anchor}(i)$ and $T_{test}(i)$ denote the encoding time using original VTM and the proposed fast algorithm under $i$th QP with $\{$22, 27, 32, 37$\}$. Additional tests are conducted in VTM 15.0 to validate the extensibility of the proposed method.

\begin{table}[t]
  \centering
  \caption{Threshold setup in the post-processing procedure}
    \begin{tabular}{c|c|c|c}
    \hline
    configuration & $\rm T_{Intra}$ & $\rm T_{QT}$ & $\rm T_{MT}$ \bigstrut\\
    \hline 
    \textit{Our: ``C1"} & 0.9 & $0.1\sim0.0$ & $0.05\sim0.0$ \bigstrut\\
    \hline
    \textit{Our: ``C2"} & 0.8 & $0.2\sim0.1$ & $0.15\sim0.1$ \bigstrut\\
    \hline
    \textit{Our: ``C3"} & 0.7 & $0.3\sim0.2$ & $0.25\sim0.2$ \bigstrut\\
    \hline
    \end{tabular}%
  \label{Thres}%
\end{table}%
\begin{table*}[htbp]
  \centering
  \caption{Performance of the fast partition algorithm in VTM 7.0 compared to the state-of-the-art works}
    \begin{tabular}{c|c|c|c|c|c|c|c|c|c|c}
    \hline
    \multirow{2}[4]{*}{Sequences} & \multicolumn{2}{c|}{TCSVT-21\cite{2021TCSVTcfg}} & \multicolumn{2}{c|}{TCSVT-22\cite{2022TCSVTHGFCN}} & \multicolumn{2}{c|}{\textit{Our: ``C1"}} & \multicolumn{2}{c|}{\textit{Our: ``C2"}} & \multicolumn{2}{c}{\textit{Our: ``C3"}} \bigstrut\\
\cline{2-11}          & {\makecell[c] {BD-BR \\ (\%)}} &  {\makecell[c] {$\Delta \rm{T}$ \\ (\%)}} & {\makecell[c] {BD-BR \\ (\%)}} & {\makecell[c] {$\Delta \rm{T}$ \\ (\%)}} & {\makecell[c] {BD-BR \\ (\%)}} & {\makecell[c] {$\Delta \rm{T}$ \\ (\%)}} & {\makecell[c] {BD-BR \\ (\%)}} & {\makecell[c] {$\Delta \rm{T}$ \\ (\%)}} & {\makecell[c] {BD-BR \\ (\%)}} & {\makecell[c] {$\Delta \rm{T}$ \\ (\%)}} \bigstrut\\
    \hline
    Class A1 & 1.05  & 53.40 & 1.37  & 56.52 & 0.50  & 48.89 & 1.05  & 58.75 & 1.89  & 63.97 \bigstrut\\
    \hline
    Class A2 & 1.11  & 52.63 & 1.26  & 62.63 & 0.46  & 48.23 & 1.11  & 61.07 & 2.00  & 68.82 \bigstrut\\
    \hline
    Class B & 1.40  & 58.26 & 1.09  & 51.11 & 0.60  & 49.53 & 1.31  & 64.75 & 2.35  & 73.33 \bigstrut\\
    \hline
    Class C & 1.65  & 51.99 & 0.95  & 41.27 & 0.61  & 38.54 & 1.37  & 55.88 & 2.50  & 67.06 \bigstrut\\
    \hline
    Class D & 1.24  & 49.92 & 1.00  & 38.00 & 0.53  & 36.79 & 1.26  & 55.09 & 2.16  & 64.58 \bigstrut\\
    \hline
    Class E & 2.07  & 58.45 & 1.32  & 47.54 & 0.89  & 47.98 & 1.82  & 63.97 & 3.11  & 73.75 \bigstrut\\
    \hline
    \textbf{Average} & \textbf{1.42} & \textbf{54.20} & \textbf{1.14} & \textbf{48.76} & \textbf{0.60} & \textbf{44.74} & \textbf{1.32} & \textbf{59.95} & \textbf{2.33} & \textbf{68.76} \bigstrut\\
    \hline
    \end{tabular}%
  \label{main}%
\end{table*}%

\begin{figure}[ht]
  \centering
  \includegraphics[width=8cm,height=6cm]{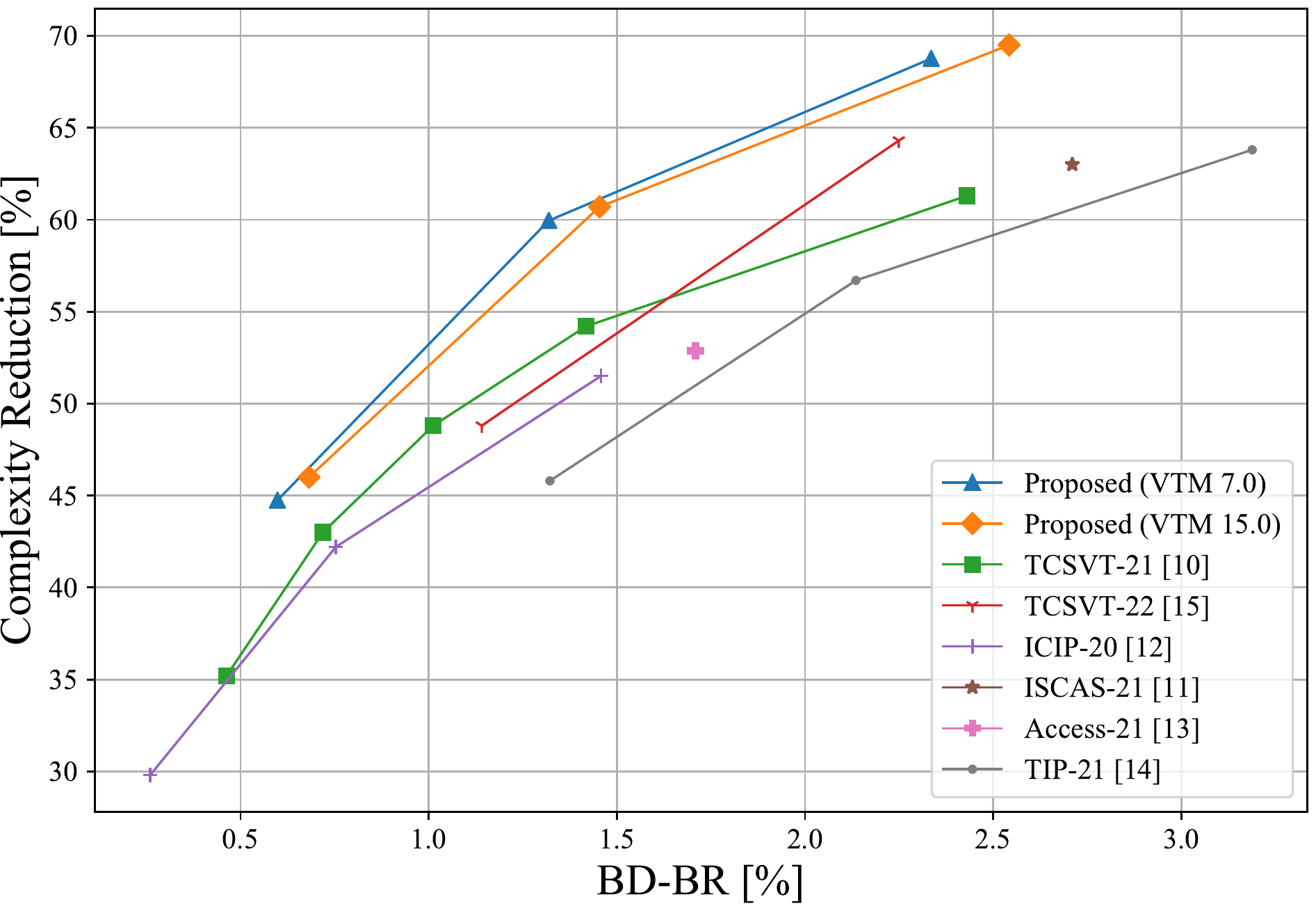}
  \caption{BD-BR loss and complexity reduction for the proposed method and comparison with the related works.}\label{curves}
\end{figure}

\subsection{Performance}
The performance tested in VTM 7.0 is shown in Table \ref{main}. The encoding complexity is reduced by 44.74\% to 68.76\% while BD-BR loss is only 0.60\% to 2.33\%. Compared with the work of Saldanha \textit{et al.} \cite{2021TCSVTcfg}, our method achieves better performance on all sequences evaluated. The experimental results of Class A1 demonstrate that our solution \textit{``C2"} surpasses the work of Wu \textit{et al.} \cite{2022TCSVTHGFCN}, saving 2.23\% more encoding time with 0.32\% less BD-BR loss. 

To further visualize the performance of the proposed method, a performance comparison curve is plotted in Fig.\ref{curves}, including several significant investigations \cite{2021TCSVTcfg, 2021ISCASSVM, 2020ICIPCNN, zhang2021fast, 2021TIPDeepQTMT, 2022TCSVTHGFCN}. Our method clearly realizes a better compromise between BD-BR loss and complexity reduction. The very similar performance in VTM 7.0 and VTM 15.0 is due to the fact that the QTMT block partition structure is unchanged in the version update. However, many new tools are added in VTM 15.0, and the overall encoding complexity is higher, so the performance in VTM 7.0 is slightly better.

\begin{table}[t]
  \centering
  \caption{Ablation results of the QP fusion network}
    \begin{tabular}{c|c|c|c|c}
    \hline
    \multirow{2}[4]{*}{Ablation} & 
    \multicolumn{2}{c|}{Number of Components} & 
    \multicolumn{2}{c}{Performance} \bigstrut\\
\cline{2-5}          & 
\multicolumn{1}{c|}{\makecell[c] {QP Fusion\\Operator}} & 
\multicolumn{1}{c|}{\makecell[c] {Conv. \\ 3×3}} & 
\multicolumn{1}{c|}{\makecell[c] {BD-BR \\ (\%)}} & 
\multicolumn{1}{c} {\makecell[c] {$\Delta \rm{T}$ \\ (\%)}} \bigstrut\\
    \hline
    \textit{S1} & 0     & 0     & 1.37  & 56.62 \bigstrut\\
    \hline
    \textit{S2} & 0     & 3     & 1.39  & 56.73 \bigstrut\\
    \hline
    \textit{S3} & 1     & 3     & 1.35  & 57.82 \bigstrut\\
    \hline
    \textit{Proposed}  & 3     & 3     & \textbf{1.31}  & \textbf{58.57} \bigstrut\\
    \hline
    \end{tabular}%
  \label{ablation}%
\end{table}%

\subsection{Ablation Experiment}
To demonstrate the effectiveness of the QP fusion network, the redesigned U-Net and the post-processing strategy are maintained the same while changing the structure of the QP fusion network. The final structure (\textit{Proposed}) is shown in Fig.\ref{framework}. The other three structures (\textit{S1}, \textit{S2} and \textit{S3}) for comparison and the average performance of Class B, C, and D are shown in Table \ref{ablation}. It can be observed that the proposed structure has the best performance. \textit{S1} is utilized as the test anchor and has no QP fusion operators or convolution layers. The number of training parameters is the same in \textit{Proposed} and \textit{S2}, excluding the influence of network complexity on performance. The result of the comparison between \textit{Proposed} and \textit{S3} with just one QP fusion operator demonstrates that additional QP fusion operators lead to superior performance. However, more cascaded QP fusion operators would instead give rise to over-complicated computations, so the three-layer cascade structure is eventually chosen.

\section{Conclusion}
\label{sec:Conclusion}

In this paper, we propose a CNN-based approach to accelerate the decision process of QTMT partition structure for VVC intra coding. Firstly, we construct the backbone network by improving U-Net. Then we design a QP fusion network and a refined post-processing strategy to boost performance further. With a BD-BR loss of 0.60\% $\sim$ 2.33\%, the proposed fast algorithm saves 44.74\% $\sim$ 68.76\% of encoding time. In future work, we plan to extend the fast partition algorithm to inter encoding and complete the hardware implementation.

\bibliographystyle{IEEEtran}
\bibliography{main}

\end{document}